\documentclass[aip,reprint]{revtex4-1}
\usepackage{graphicx}
\usepackage{ulem}
\usepackage{amssymb}
\usepackage{amsmath}
\usepackage{wasysym}
\usepackage{relsize}
\begin{document}
%\draft                    

%\flushbottom
%\twocolumn[
%\hsize\textwidth\columnwidth\hsize\csname @twocolumnfalse\endcsname

\newcommand{\be}{\begin{equation}}
\newcommand{\ee}{\end{equation}}
\newcommand{\bea}{\begin{eqnarray}}
\newcommand{\eea}{\end{eqnarray}}
\newcommand{\Tbar}{{\bar{T}}}
\newcommand{\En}{{\cal E}}
\newcommand{\K}{{\cal K}}
\newcommand{\U}{{\cal U}}
\newcommand{\GC}{{\cal \tt G}}
\newcommand{\Lop}{{\cal L}}
\newcommand{\DB}[1]{\marginpar{\footnotesize DB: #1}}
\newcommand{\q}{\vec{q}}
\newcommand{\kt}{\tilde{k}}
\newcommand{\Lopn}{\tilde{\Lop}}
\newcommand{\noi}{\noindent}
\newcommand{\ovn}{\bar{n}}
\newcommand{\ovx}{\bar{x}}
\newcommand{\ovE}{\bar{E}}
\newcommand{\ovV}{\bar{V}}
\newcommand{\ovU}{\bar{U}}
\newcommand{\ovJ}{\bar{J}}
\newcommand{\calE}{{\cal E}}
\newcommand{\ovphi}{\bar{\phi}}
\newcommand{\zt}{\tilde{z}}
\newcommand{\ttl}{\tilde{\theta}}
\newcommand{\nuv}{\rm v}
\newcommand{\ds}{\Delta s}
\newcommand{\fn}{{\small {\rm  FN}}}
\newcommand{\cc}{{\cal C}}
\newcommand{\tth}{\tilde{\theta}}
\newcommand{\cb}{{\cal B}}
\newcommand{\cg}{{\cal G}}
\newcommand\norm[1]{\left\lVert#1\right\rVert}
\title{The cosine law of field enhancement factor variation: generic emitter shapes}
 
%\vskip 0.3 in

\author{Debabrata Biswas}
\affiliation{
Bhabha Atomic Research Centre,
Mumbai 400 085, INDIA}
\affiliation{Homi Bhabha National Institute, Mumbai 400 094, INDIA}
\author{Gaurav Singh}
\affiliation{
Bhabha Atomic Research Centre,
Mumbai 400 085, INDIA}
\affiliation{Homi Bhabha National Institute, Mumbai 400 094, INDIA}
\author{Rajasree Ramachandran}
\affiliation{
Bhabha Atomic Research Centre,
Mumbai 400 085, INDIA}

%\pacs{85.45.-w}{}
%\pacs{03.65.Sq}{}
%\pacs{03.65.Xp}{}
%\pacs{52.59.Sa}{}

\begin{abstract}
  The cosine law of field enhancement factor variation was recently derived for a hemi-ellipsoidal emitter
  and numerically established for other smooth emitter shapes (Biswas et al, Ultramicroscopy, 185, 1 (2018)).
  An analytical derivation is provided here for general smooth vertical emitter shapes aligned in the
  direction of the asymptotic electrostatic field. The law is found to hold in the neighbourhood of the emitter
  apex from where field emission pre-dominantly occurs.
\end{abstract}

%\maketitle

%\pacs{85.45.-w}{}
%\pacs{03.65.Sq}{}
%\pacs{03.65.Xp}{}
%\pacs{52.59.Sa}{}

%\date{\today}
%\vskip 0.2 in
%\centerline{\bf Abstract}

%\vskip 0.25 in

%\pacs{85.35.-p, 03.65.Sq, 52.59.Sa}

\maketitle

%]

%\newpage
%\noindent

%\section{Introduction}
%\label{sec:Introduction}

\section{Introduction}
\label{sec:intro}

Field emitters finds application in various vacuum nanoelectronics devices where a cold, bright source of electrons
is required. They generally involve ultrasharp emitting tips with apex radius of curvature in the nanometer regime.
This leads to enhancement of the local field on the emitter surface so that electric fields of the order of $\text{V/nm}$
are generated even at moderate applied voltages.
 
A measure of local field enhancement is the {\it apex field enhancement factor}. It refers to
the ratio of the magnitude of the local electric field at
an emitter apex to the asymptotic electric field away from the
cathode plane \cite{ev2002,forbes2003,Read_Bowring,podenok,pogorelov,db_fef,db_rudra}.
While this is an important quantity in field emission theory \cite{FN,Nordheim,murphy,forbes,forbes_deane,jensen_ency}
and a topic of considerable research, a calculation
of the net emission current also requires knowledge about the variation of the enhancement factor
close to the emitter apex \cite{dyke,podenok,Read_Bowring,db_ultram,db_distrib}.
In Ref. [\onlinecite{db_ultram}], it was shown that the field enhancement factor
$\gamma$ at any point close to the apex of an axially symmetric emitter, is related to
the apex field enhancement factor $\gamma_a$ by

\be
\gamma = \gamma_a \cos{\tth}  \label{eq:coslaw}
\ee

\noi
where

\be
\cos{\tth} = \frac{z/h}{\sqrt{(z/h)^2 + (\rho/R_a)^2}},  \label{eq:cosdef}
\ee

\noi
$h$ is the height of the emitter, $R_a$ its apex radius of curvature and
$(\rho,z)$ is a point close to the emitter apex ($\rho = 0, z = h$). The underlying assumption
is that the emitter is aligned along the asymptotic electrostatic field $E_0 \hat{z}$.
Eq.~\ref{eq:coslaw} was established analytically \cite{db_ultram} for a hemi-ellipsoid
emitter and numerically found to be true for other emitter shapes including a cone and
a cylindrical post with a parabolic cap.

In the following, we shall establish Eq.~\ref{eq:coslaw} analytically for general emitter shapes
starting from the line charge model \cite{pogorelov,harris15,jap16}.

\section{A model for general emitter shapes}

\begin{figure}[hbt]
  \begin{center}
    \vskip -1.9cm
%\hskip -1.8cm
\hspace*{-1.0cm}\includegraphics[scale=0.35,angle=0]{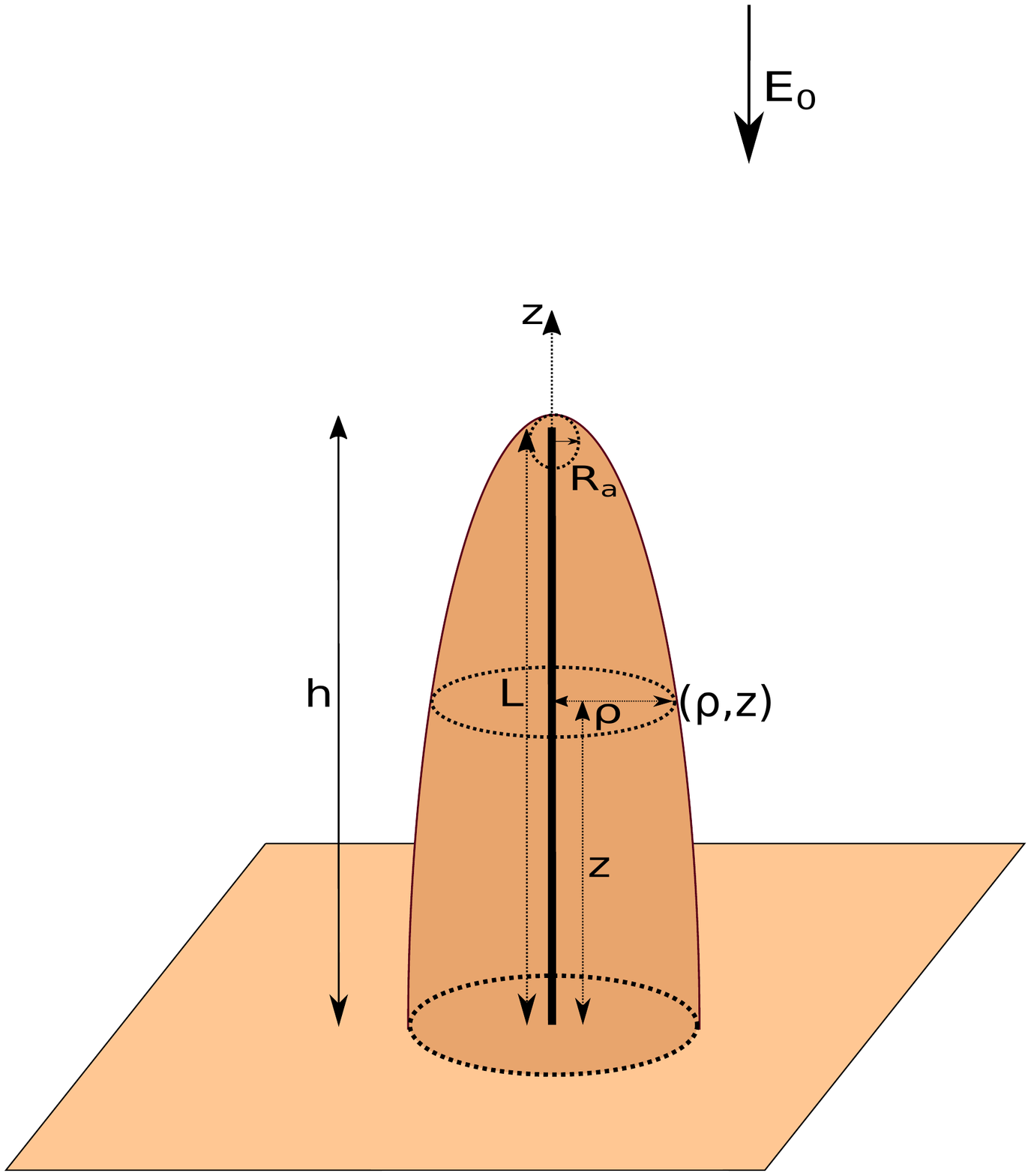}
\vskip -1.0 cm
\caption{An emitter mounted on a metallic cathode plate in the presence of an asymptotic
  field $-E_0 \hat{z}$. The emitter is equivalently modeled as a line charge distribution (bold line).}
\label{fig:model}
\end{center}
\end{figure}

Consider an emitter mounted on an infinite metallic cathode plate, aligned in the direction of the
asymptotic electric field $-E_0 \hat{z}$ (see Fig.~\ref{fig:model}). Alternately, consider a parallel-plate geometry
with the two plates separated by a distance $D$, the emitter mounted on a grounded cathode
and the anode at a positive potential $V$. If $D$ is large compared to the height $h$ of emitter, the
anode has negligible effect on the emitter apex and the asymptotic field $E_0 \simeq V/D$.

The termination of the field lines on the cathode surface gives rise to a surface charge density
$\sigma(\rho,z)$, which in turn can be projected on the emitter axis as a line charge density, $\Lambda(z)$.
For the hemi-ellipsoid in an asymptotic field $E_0$, it is known that $\Lambda(z)  = \lambda z$ while
in general, the surface and line charge densities are related as \cite{jap16}

\be
\Lambda(z) = 2\pi \rho(z) \sqrt{1 + (d\rho/dz)^2} \sigma(z)
\ee

\noi
where $\rho = \rho(z)$ defines the surface of the axially symmetric emitter. In general, $\Lambda(z)$
is expected to be a smooth nonlinear function of $z$ with the nonlinear terms depending on the excursion
from the hemi-ellipsoidal shape. Note that close to the emitter base, the field lines are expected to
terminate at the cathode plate so that $\Lambda(z) \rightarrow 0$ for $z \rightarrow 0$.
A convenient form for the line charge density of such a vertically aligned emitter is thus
$\Lambda(z) = z f(z)$ where $f(z)$ depends on the emitter shape and is otherwise unknown.
We refer to this as the line charge model for a generic emitter shape.

The potential at any point ($\rho,z$) due to a vertical line charge placed on a grounded conducting
plane can be expressed as

\be
\begin{split}
V(\rho,z) = & \frac{1}{4\pi\epsilon_0}\Big[ \int_0^L \frac{\Lambda(s)}{\big[\rho^2 + (z - s)^2\big]^{1/2}} ds ~
  - \\
  &  \int_0^L \frac{\Lambda(s)}{\big[\rho^2 + (z + s)^2\big]^{1/2}} ds \Big] + E_0 z \label{eq:pot}
\end{split}
\ee

\noi
where $L$ is the extent of the line charge distribution. Note that the second integral arises
from the image of the line charge distribution. The zero-potential contour
corresponds to the surface of the desired emitter shape so that
the parameters defining the line charge distribution
including its extent $L$, can, in principle be calculated by imposing the requirement that
the potential should vanish on the surface of the emitter.

Starting with Eq.~\ref{eq:pot}, we shall investigate the cosine law of Eq.~\ref{eq:coslaw}
for a general sharp emitter with $h >> R_a$.

\section{The Electric Field Components on the Surface and the cosine law}

The surface of a vertically aligned emitter $z = z(\rho)$  can be expressed in the neighbourhood
of the apex by Taylor expanding $z$ at $\rho = 0$. Since $h = z(0)$ and $dz/d\rho = 0$
for a vertically aligned emitter,

\be
z \simeq h - \frac{\rho^2}{2 R_a}
\ee

\noi
in the immediate vicinity of the apex from where field emission occurs. The electric field lines
are normal to this surface and thus in the direction

\be
\hat{n} = \frac{1}{\sqrt{1 + (\rho/R_a)^2}} (\frac{\rho}{R_a} \hat{\rho},\hat{z}).
\ee

\noi
It is thus necessary to establish that $|\vec{E}| = \vec{E}.\hat{n} = E_a \cos\tth $ in order that the
cosine law is valid for a general $\Lambda(z)$ where

\be
\vec{E} = E_\rho ~\hat{\rho} + E_z ~\hat{z}  =
-\Big( \frac{\partial V}{\partial \rho} \hat{\rho} + \frac{\partial V}{\partial z} \hat{z} \Big)
\ee

\noi
and $E_a$ is the field at the apex. Alternately, it can be shown that $|\vec{E}| =
\sqrt{E_\rho^2 + E_z^2} = E_a \cos\tth$. We shall pursue both approaches in deriving the cosine
law for a sharp emitter.

\subsection{Linear line charge density}

We shall first deal with linear line charge density $\Lambda (z) = \lambda z$ where $\lambda$ is a constant which can be evaluated using $V(0,h) = 0$. Thus,

\be
\lambda = - \frac{4\pi\epsilon_0 E_0 h}{h\ln\Big(\frac{h+L}{h-L}\Big) - 2L} \label{eq:lam}
\ee

\noi
where $L$ is related to the height $h$ and apex radius of curvature $R_a$ through the relation
$(h^2 - L^2)/h = R_a$ which holds for linear as well as nonlinear charge distributions \cite{db_fef}.
The field components can be calculated by differentiating
Eq.~\ref{eq:pot} and the integrals are easy to evaluate when the line charge density is
linear. The methods employed are however sufficiently general to be of
use in the nonlinear case as well \cite{db_fef}.

The $\hat{\rho}$ component of the electric field, -$\partial V/\partial \rho$ can be evaluated to yield

\be
\begin{split}
E_\rho = \frac{\lambda}{4\pi\epsilon_0} \frac{1}{\rho} \Bigg[ \frac{\rho^2 + z(z+ L)}{\sqrt{\rho^2 + (z + L)^2}} - &
  \frac{\rho^2 + z(z - L)}{\sqrt{\rho^2 + (z - L)^2}} \Bigg]
\end{split}
\ee

\noi
while the $\hat{z}$ component of the field is

\be
\begin{split}
E_z = & \frac{\lambda}{4\pi\epsilon_0}  \Bigg[ \frac{L}{\sqrt{\rho^2 + (z + L)^2}} + \frac{L}{\sqrt{\rho^2 + (z - L)^2}} ~+  \\
&  \ln\Big\{ \frac{\sqrt{\rho^2 + (z+L)^2} - (z+L)}{\sqrt{\rho^2 + (z-L)^2} - (z-L)} \Big\}\Bigg] - E_0 . \label{eq:Ez0}
\end{split}
\ee

\noi
For small values of $\rho$ near the tip,

\be
\begin{split}
E_\rho \simeq  \frac{\lambda}{4\pi\epsilon_0} \frac{1}{\rho} & \Bigg[ \big(z + \frac{\rho^2}{z+L}\big)\big(1 -
  \frac{\rho^2}{2(z+L)^2}\big) ~- \\
  & \big(z + \frac{\rho^2}{z-L}\big)\big(1 -
  \frac{\rho^2}{2(z-L)^2}\big) \Bigg]
\end{split}
\ee

\noi
which, on further simplification yields

\be
E_\rho \simeq \frac{\lambda \rho}{4\pi\epsilon_0} \Bigg[ \frac{2z^2 L}{(z^2 - L^2)^2} - \frac{2L}{(z^2 - L^2)} \Bigg].
\ee

\noi
Since $z \sim h$ and $z^2 - L^2 \sim h R_a$, the first term dominates for a sharp emitter. Thus,

\be
E_\rho \simeq \frac{\lambda}{4\pi\epsilon_0} \frac{2zL}{(z^2 - L^2)} \frac{z\rho}{(z^2 - L^2)} \label{eq:Erho}
\ee

\noi
for small $\rho$.

The $\hat{z}$ component  of the field can be similarly simplified. Neglecting the logarithmic term and $E_0$ for a sharp emitter, Eq.~\ref{eq:Ez0}
can be expressed as

\be
\begin{split}
E_z \simeq \frac{\lambda}{4\pi\epsilon_0} & \Bigg[ \frac{L}{z-L}\big(1 - \frac{\rho^2}{2(z-L)^2} \big) + \\
& \frac{L}{L+z}\big(1 - \frac{\rho^2}{2(z+L)^2} \big) \Bigg]
\end{split}
\ee

\noi
which can be further simplified as

\be
E_z \simeq \frac{\lambda}{4\pi\epsilon_0} \frac{2zL}{(z^2 - L^2)}  \Bigg[1 - \frac{\rho^2(z^2 + 3L^2)}
  {2(z^2 - L^2)^2} \Bigg]. \label{eq:Ez}
\ee

We are now in a position to establish the cosine law using $\vec{E}.\hat{n}$ or $|E| = \sqrt{E_\rho^2 + E_z^2}$.
Note that on the surface of the emitter close to the apex, $\rho$ and $z$ are related by $z = h - \rho^2/(2R_a^2)$.
Also, $L = \sqrt{h(h-R_a)} \simeq h$ and

\be
\frac{2zL}{z^2 - L^2} \simeq \frac{2h}{R_a} (1 + \frac{\rho^2}{R_a^2})
\ee

\noi
so that

\be
\begin{split}
E_\rho & \simeq  \frac{\lambda}{4\pi\epsilon_0} \frac{2h}{R_a} ( 1 + \frac{\rho^2}{R_a^2}) \frac{\rho}{R_a} \\
E_z & \simeq \frac{\lambda}{4\pi\epsilon_0} \frac{2h}{R_a} ( 1 + \frac{\rho^2}{R_a^2}) \Big[1 - 2 \frac{\rho^2}{R_a^2} \Big]
\end{split}
\ee

Thus, close to the apex, using the relation between $z$ and $\rho$ on the emitter surface,

\be
%\begin{split}
  \vec{E}.\hat{n} =  \frac{\lambda}{4\pi\epsilon_0}   \frac{2h}{R_a}   \frac{(1 +  \frac{\rho^2}{R_a^2})}{\sqrt{(1 + \frac{\rho^2}{R_a^2})}}  \Bigg [\frac{\rho^2}{R_a^2}  + 1 - \frac{2\rho^2}{R_a^2} \Bigg].
%\end{split}
\ee

\noi
Since, we are interested in the variation close to the apex ($\rho$ small),

\bea
\vec{E}.\hat{n} & = & \gamma_a E_0 \frac{(1 - \rho^4/R_a^2)}{\sqrt{1 + \rho^2/R_a^2}} \\
& \simeq & \frac{\gamma_a E_0}{\sqrt{1 + \rho^2/R_a^2}} \simeq \gamma_a E_0 \cos\tth
\eea

\noi
where we have used $(\lambda/4\pi\epsilon_0)(2h/R_a) = E_a = \gamma_a E_0$. Alternately,

\be
\begin{split}
  |E| = & \sqrt{E_\rho^2 + E_z^2} \\
  & \simeq \frac{\lambda}{4\pi\epsilon_0} \frac{2zL}{(z^2 - L^2)} \Bigg[ \frac{z^2\rho^2}{(z^2 - L^2)^2} + 1 \\
    & - \frac{\rho^2}{(z^2 - L^2)^2} (z^2 + 3L^2) \Bigg]^{1/2} \\
  & = \frac{\lambda}{4\pi\epsilon_0} \frac{2zL}{(z^2 - L^2)}\Bigg[ 1 - \frac{3L^2\rho^2}{(z^2 - L^2)^2} \Bigg]^{1/2}
\end{split}
\ee

\noi
which approximates to

\be
\begin{split}
|E| & \simeq   \frac{\lambda}{4\pi\epsilon_0} \frac{2h}{R_a}\Big(1 + \frac{\rho^2}{R_a^2}\Big) \Big(1 -
\frac{3\rho^2}{2R_a^2} (1 + 2 \frac{\rho^2}{R_a^2}) \Big) \\
& \simeq \frac{\lambda}{4\pi\epsilon_0}  \frac{2h}{R_a} \Big(1 + \frac{\rho^2}{R_a^2}\Big) \Big(1 - \frac{3\rho^2}{2R_a^2}\Big).
\end{split}
\ee

\noi
This can be further simplified to yield 

\be
\begin{split}
|E|  & \simeq \frac{\lambda}{4\pi\epsilon_0}
  \frac{2h}{R_a}  \Big( 1 - \frac{1}{2} \frac{\rho^2}{R_a^2} \Big) \\
& \simeq \frac{\gamma_a E_0}{\sqrt{1 + \rho^2/R_a^2}} \simeq \gamma_a E_0 \cos\tth.
\end{split}
\ee

\noi
Thus, the cosine law of field enhancement factor has been shown to hold close to the apex
from where emission predominantly occurs.

\subsection{Nonlinear Line Charge Density}

As discussed earlier, the line charge density may be assumed to be of the form $\Lambda(z) = z f(z)$
without any loss of generality. As in the linear case, we shall first derive expressions for
$E_\rho$ and $E_z$ close to the apex ($\rho$ small).

The $E_\rho$ component can be expressed as

\be
\begin{split}
  E_\rho = -\frac{\partial V}{\partial \rho} & = -\frac{\rho}{4\pi\epsilon_0} \Bigg[ \int_0^L \frac{s f(s)}{[\rho^2 + (z+s)^2]^{3/2}}ds - \\
      & \int_0^L \frac{s f(s)}{[\rho^2 + (z-s)^2]^{3/2}}ds \Bigg] 
\end{split}
\ee

\noi
For $\rho$ small,

\be
\begin{split}
  E_\rho =  &  -\frac{\rho}{4\pi\epsilon_0} \Bigg[ \int_0^L ds \frac{s f(s)}{(z+s)^3} \Big[ 1 - \frac{3\rho^2}{2(z+s)^2} \Big] - \\
    & \int_0^L ds \frac{s f(s)}{(z-s)^3} \Big[ 1 - \frac{3\rho^2}{2(z-s)^2} \Big] \Bigg]. \label{eq:Erhoexpansion}
\end{split}
\ee

\noi
Note that as before $z \simeq h$. The expansion in the second integral is justified since
$z^2 - L^2 = hR_a$ implies $z - L \simeq R_a/2$. Since $z-s$ is bounded from below by $z - L$,
the expansion holds so long as $\rho < R_a/2$.

We shall first deal with the leading terms in the integral:

\be
\begin{split}
 \int_0^L   \Bigg[ & \frac{s f(s)}{(z+s)^3} -  \frac{s f(s)}{(z-s)^3} \Bigg] ds \\
 =   &  - \Bigg[ f(s) \frac{2s^3}{(z^2 - s^2)^2} \Bigg]_0^L +  \int_0^L f'(s) \frac{2s^3}{(z^2 - s^2)^2} ds \\
 = & - f(L) \frac{2L^3}{(z^2 - L^2)^2} \Bigg[ 1 - \int_0^L ds~ \frac{f'(s)}{f(L)} \frac{s^3/(z^2 - s^2)^2}{L^3/(z^2 - L^2)^2} \Bigg]
\end{split}
\ee

\noi
Thus,

\be
E_\rho = f(L) \frac{2L^2}{(z^2 - L^2)} \frac{L \rho}{z^2 - L^2} \Big[1 - \cc_0 \Big]
\ee

\noi
where

\be
\cc_0 = \int_0^L ds~ \frac{f'(s)}{f(L)} \frac{s^3/(z^2 - s^2)^2}{L^3/(z^2 - L^2)^2}.
\ee

\noi
Since the charge distribution is well behaved for a smooth emitter and can be expressed as a polynomial function of
degree $n$ (for cases of interest here, $n \leq 5$), $f(s)$ obeys Bernstein's inequality \cite{ineq}

\be
|f'(x)| \leq \frac{n}{(1 - x^2)^{1/2}} \norm{f}  \label{eq:bern}
\ee

\noi
where $x \in [-1,1]$ and $\norm{f}$ denotes the maximum value of $f$ in this interval. With
$x = s/h$ and applying the inequality, it can be shown that $\cc_0 \sim (z^2 - L^2)^{1/2}$ is
vanishingly small for sharp-tipped emitters since $z \simeq h$. 
The dominant contribution to $E_\rho$ is

\be
E_\rho \simeq \frac{f(L)}{4\pi\epsilon_0} \frac{2z L}{(z^2 - L^2)}\frac{z\rho}{(z^2 - L^2)}  \label{eq:Erhofin}
\ee

\noi
Note that the $\rho^2$ correction terms in the integrand of Eq.~\ref{eq:Erhoexpansion} can be neglected
for small $\rho$. Thus Eq.~\ref{eq:Erhofin} is the final expression for $E_\rho$ close to the apex.

The $z$ component of the electric field can be similarly evaluated. Thus,

\be
\begin{split}
  E_z = -\frac{\partial V}{\partial z} = & -\frac{1}{4\pi\epsilon_0} \Bigg[
    - \int_0^L ds~ \frac{(z-s) sf(s)}{[\rho^2 + (z-s)^2]^{3/2}} \\
    & +  \int_0^L ds~ \frac{(z+s) sf(s)}{[\rho^2 + (z+s)^2]^{3/2}}\Bigg] + E_0
\end{split}
\ee

\noi
For $\rho$ small ($\rho < R_a/2$ as mentioned earlier),

\be
\begin{split}
  E_z =  & -\frac{1}{4\pi\epsilon_0} \Bigg[
     \int_0^L ds~sf(s)\Big\{\frac{1}{(z+s)^2} -  \frac{1}{(z-s)^2}  \Big\} - \\
    & \frac{3}{2}\rho^2 \int_0^L ds~sf(s)\Big\{\frac{1}{(z+s)^4} -  \frac{1}{(z-s)^4}  \Big\} \Bigg] - E_0 
\end{split}
\ee

\noi
We shall first consider the integral

\be
\begin{split}
  \int_0^L & ds ~s f(s) \Big\{ \frac{1}{(z+s)^2} - \frac{1}{(z-s)^2} \Big\}  \\
  = & \Bigg[ \Big\{-\frac{2zs}{(z^2 - s^2)} + \ln\frac{z+s}{z-s} \Big\} f(s) \Bigg]_0^L \\
  & + \int_0^L ds f'(s) \frac{2zs}{z^2 - s^2} - \int_0^L f'(s) \ln\frac{z+s}{z-s} \\
  = & -\frac{2zL}{z^2 - L^2} f(L) \Big[ 1 - \cc_1 \Big]  + \ln\frac{z+s}{z-s} \Big[ 1 - \cc_2 \Big] \\
 \simeq & -\frac{2zL}{z^2 - L^2} f(L) \Big[ 1 - \cc_1 \Big] \label{eq:int1}
\end{split}
\ee

\noi
where

\bea
\cc_1 & = &  \int_0^L \frac{f'(s)}{f(L)} \frac{s/(z^2 - s^2)}{L/(z^2 - L^2)} ds \\
\cc_2  & =  & \int_0^L \frac{f'(s)}{f(L)}\frac{\ln\Big(\frac{z+s}{z-s}\Big)}{\ln\Big(\frac{z+L}{z-L}\Big)} ds.
\eea

\noi
Using Bernstein's inequality, $\cc_1$ is negligible for a sharp emitter (${\cal{O}}\big(z^2 - L^2)^{1/2} \big)$ but $\cc_2$ cannot be neglected due to the logarithmic
dependence. However, $\frac{2zL}{z^2 - L^2} >>  \ln\frac{z+L}{z-L}$ for a sharp emitter so that the
last line of Eq.~\ref{eq:int1} follows.

Consider now the integral

\be
\int_0^L ds~\Big\{ \frac{sf(s)}{(z+s)^4} - \frac{sf(s)}{(z-s)^4} \Big\}.
\ee

\noi
Using the methods adopted earlier, this reduces to

\be
\begin{split}
  \Big[ - \frac{8}{3} & f(s) \frac{s^3 z}{(z^2 - s^2)^3} \Big]_0^L  + \\
  & \int_0^L f'(s) \frac{8}{3} \frac{s^3 z}{(z^2 - s^2)^3} ds \\
  = & -\frac{8}{3} f(L) \frac{L^3z}{(z^2 - L^2)^3} \Big[1 - \cc_0 \Big]
\end{split}
\ee

\noi
It follows from Bernstein's inequality that $\cc_0 \sim (z^2 - L^2)^{1/2}$ which is vanishingly small. Thus,

\be
E_z \simeq \frac{f(L)}{4\pi\epsilon_0} \frac{2zL}{z^2 - L^2} \Bigg[ 1 -
  \frac{\rho^2}{2} \frac{4L^2}{(z^2 - L^2)^2} \Bigg]. \label{eq:Ezfin}
\ee

\noi
Since $z \simeq L$ close to the emitter apex, both $E_\rho$ and $E_z$ (Eqns. \ref{eq:Erhofin} and \ref{eq:Ezfin} respectively) have the same
form as in the linear case derived earlier (Eqns. \ref{eq:Erho} and \ref{eq:Ez} respectively) with $f(L)$ replacing $\lambda$.
Note also that the linearly varying line charge density is a special case of the nonlinear density with $f'(s) = 0$
so that the correction terms $\cc_0,\cc_1,\cc_2$ are identically equal to zero instead of being merely small.
Furthermore, as shown in Ref. [\onlinecite{db_fef}],

\be
\frac{f(L)}{4\pi\epsilon_0} \frac{2zL}{z^2 - L^2} = E_0 \gamma_a = E_a
\ee

\noi
so that the cosine law follows as in the linear case.

\section{Discussion and Conclusions}

The local field at the emitter apex and its neighbourhood is a very important factor in field emission calculations.
Using the line-charge formalism applicable for a vertically aligned axially symmetric emitter in a diode configuration,
we have established the cosine law of field enhancement factor variation close to the emitter apex. The general result for
a nonlinear line charge is derived in the same spirit as for a linear line charge distribution
where it is known that the cosine law is exact \cite{pogorelov,db_ultram} and in fact valid further away from the emitter apex.

The results presented here are valid for a single emitter when the anode is sufficiently far away for the line
charge and its image to have negligible effect on the anode. This is generally true when the anode-cathode plate
distance is more than than three times the height of the emitter. Since this condition may hold in practical devices, the
anode proximity effect is not a major impediment to the cosine law.

Isolated emitters, while of immense interest in microscopy, are not the norm in cathodes. Bright electron sources
have large area field emitters (LAFE) with several emitters in an array or randomly placed. It is not too difficult to
establish that the cosine law is valid here as well, at least for emitters that are well separated and contribute
significantly to the LAFE current. Thus, the cosine law is applicable in general situations and may be used to
evaluate the field emitter current.

\section{Acknowledgment}

The authors acknowledge useful discussions with Dr. Raghwendra Kumar and Rashbihari Rudra.

\vskip -0.25 in
$\;$\\
\section{References} 

%\begin{references}

\end{document}